Biological
Cybernetics

# Developmental time windows for axon growth influence neuronal network topology

**Sol Lim · Marcus Kaiser**

**Abstract**  Early brain connectivity development consists of multiple stages: birth of neurons, their migration and the subsequent growth of axons and dendrites. Each stage occurs within a certain period of time depending on types of neurons and cortical layers. Forming synapses between neurons either by growing axons starting at similar times for all neurons (much-overlapped time windows) or at different time points (less-overlapped) may affect the topological and spatial properties of neuronal networks. Here, we explore the extreme cases of axon formation during early development, either starting at the same time for all neurons (parallel, i.e., maximally overlapped time windows) or occurring for each neuron separately one neuron after another (serial, i.e., no overlaps in time windows). For both cases, the number of potential and established synapses remained comparable. Topological and spatial properties, however, differed: Neurons that started axon growth early on in serial growth achieved higher out-degrees, higher local efficiency and longer axon lengths while neurons demonstrated more homogeneous connectivity patterns for parallel growth. Sec-

S. Lim · M. Kaiser
Department of Brain and Cognitive Sciences, Seoul National University, Seoul, Republic of Korea

S. Lim · M. Kaiser (✉)
Interdisciplinary Computing and Complex BioSystems Group (ICOS), School of Computing Science, Newcastle University, Claremont Tower, Newcastle upon Tyne NE1 7RU, UK
e-mail: m.kaiser@ncl.ac.uk

M. Kaiser
Institute of Neuroscience, Newcastle University, Newcastle upon Tyne, UK

ond, connection probability decreased more rapidly with distance between neurons for parallel growth than for serial growth. Third, bidirectional connections were more numerous for parallel growth. Finally, we tested our predictions with *C. elegans* data. Together, this indicates that time windows for axon growth influence the topological and spatial properties of neuronal networks opening up the possibility to a posteriori estimate developmental mechanisms based on network properties of a developed network.

## 1 Introduction

The formation of synapses between neurons is influenced by genetic, activity-dependent, molecular and mechanical cues in combination and also in different temporal and spatial scales (Sperry 1963; Yamamoto et al. 2002; Yu et al. 2012; Franze 2013; Scheiffele et al. 2000; Dickson 2002). To avoid abnormal functionality, finding specific target neurons is important. Many guidance mechanisms ensure the correct specification between neurons to successfully establish appropriate synapses. One of the important mechanisms for global synaptic connectivity is chemotaxis. Diffusible and membrane-bound chemical cues guide axons to find their targets (Dickson 2002; Scheiffele et al. 2000; Dickson 2002). Electrical activity also affects synaptogenesis and its reorganization (Butz et al. 2014; Butz and van Ooyen 2013). These guidance cues, however, cannot fully explain certain features of synaptic connectivity. In *C. elegans*, for example, around 40 % of connection patterns cannot be accounted for by differences in gene expression patterns (Kaufman et al.

2006; Baruch et al. 2008). Activity-dependent mechanisms are crucial for the refinement of neuronal circuits (Van Ooyen et al. 1995; Butz et al. 2009), but activity seems to have a lower influence on the early connectivity in neural systems. For example, several patterns of connectivity are preserved in knockout studies with no neurotransmitter release (Verhage et al. 2000). Short-range connectivity within less than 700 μm, or interneuron connectivity (Packer et al. 2013; Packer and Yuste 2011; Price et al. 2011) is difficult to be explained by chemical affinity guidance cues unlike long-range connectivity at the global level (Kaiser et al. 2009). Peters' rule (Braitenberg and Schüz 1998) suggests that synapse formation in brain circuitry mainly depends on the overlap of geometrical locations of specific axons and dendrites in the absence of guidance cues (Binzegger et al. 2004; van Pelt and van Ooyen 2013; McAssey et al. 2014; van Ooyen et al. 2014). In particular, the specificity of early synapse formation to develop functional neural circuits was predicted by simple factors from the overlap of axons and dendrites in hatchling frog tadpole (Li et al. 2007) and also in neocortical microcircuits (Hill et al. 2012; Packer and Yuste 2011).

Another crucial factor is the developmental time window of a neuron. Brain development occurs at different time periods depending on regions, cell types and types of development (Andersen 2003; Rakic 2002; Shaw et al. 2008). Cells are born, differentiate, migrate to certain regions of the brain and form synaptic connections influenced by the aforementioned factors. Initial overproductions of neurons and synapses are reduced after 1 year from birth, suggesting particular time periods for neurogenesis, programmed apoptosis, early synaptic pruning and synaptogenesis (Rakic et al. 1986; Purves and Lichtman 1980; Kelsch et al. 2010; Huttenlocher 1984). Time windows of development for connections between brain regions influence the topology of cortical connectivity. Kaiser and Hilgetag (Kaiser and Hilgetag 2007) and Nisbach and Kaiser (Nisbach and Kaiser 2007) showed that overlapping time windows of development could generate clusters in brain networks by having more connections between neurons with overlapping time windows of network development. Whereas these time windows operate on the population level, Varier and Kaiser (Varier and Kaiser 2011) found that neurons having similar birth times were more likely to be connected in *C. elegans*, indicating preferential synaptic connections between neurons with overlapping time windows for axon growth. Some studies have also reported preferential electrical coupling between neurons sharing genetic lineage that are likely to have similar developmental time windows in mice neocortex (Yu et al. 2009, 2012). Furthermore, non-overlapping time windows among neurons in CA3 resulted in selective synaptic connectivity forming sub-modules in hippocampus (Deguchi et al. 2011; Druckmann et al. 2014).

To investigate neuronal innervations, computational models (van Ooyen 2003, 2011) have been developed ranging from abstract models with few assumptions (Kaiser et al. 2009; Willshaw and von der Malsburg 1976; Perin et al. 2013) to more complex models simulating neurogenesis and synaptogenesis with realistic neuronal morphologies forming layers and large-scale neuronal networks in the brain (Koene et al. 2009; Godfrey et al. 2009; Hennig et al. 2009; Zubler and Douglas 2009). In previous models, however, while time windows for neural migration and synapse formation were included, it was not systematically studied how different time windows of axon growth would affect the characteristics of the brain network organization.

Here, we investigated how different time windows of axon growth would affect morphological, topological and spatial properties of short-range brain connectivity during early development. Whereas previous studies dealt with time windows that operate on the population level, our current study observes the effect of the timing of axon growth for individual neurons within a neural population. We compared two scenarios of non-overlapping (serial growth) and completely overlapping time windows (parallel growth). To study the role of time windows for axon growth in short-range connectivity, we used the approach of random axon outgrowth, randomly picking a direction, growing in a straight line and establishing a synapse when a target neuron is within certain proximity along the growth direction.

## 2 Materials and methods

Most of the assumptions of the model were endowed from the previous study (Kaiser et al. 2009).

### 2.1 Simulation

#### 2.1.1 Placement of neurons

Neurons were placed randomly in 3D space of 34 by 34 by 34 units except a neuron located at the center, which was about 4 times as many as for 2D in order to have a comparable chance for establishing a cell position without cell overlap (Kaiser et al. 2009). The number of neurons varied in the given space: 1,000, 1,400 and 1,800. The total volume of neurons relative to the embedding space ranged from 1 to 14 % depending on the cell size and the number of neurons in the simulation.

#### 2.1.2 Sizes of neurons

A neuron was simplified as a sphere where the soma and dendrites were included. The radius of a neuron varied from 0.5 to 0.9 (0.500, 0.604, 0.735, 0.900) to make the volume of the sphere increase 1, 2, 4 and 8 times in each condition. There is an upper limit for the number of incoming connections for a neuron as space around (Stepanyants et al. 2002) or on the dendrite (Kaiser et al. 2009) as well as other factors (van Ooyen et al. 2001; van Ooyen 2001) constraints the extent of synapse formation (see Sect. 2.1.5 below). The maximum values of incoming connections were varied 1, 2, 4 and 8 according to volume of a neuron because the coverage of dendritic ramifications would expand as the size of our neuron sphere increased. The radius and the number of maximum incoming connection were fixed in each condition. We assumed that the dendrites of a neuron did not grow as the size of a neuron sphere was fixed during simulations. However, we could also confirm that dendritic outgrowth, changing the size of our neurons (including soma and dendritic tree size) during development, did not affect our conclusions (Figure A9); dendrites were assumed to grow radially away from the soma based on the somatofugal growth of neurites (Samsonovich and Ascoli 2003).

### 2.1.3 Growth direction

The direction of axonal outgrowth was randomly chosen uniformly in the 3D space, and an axon grew in a straight line toward the given direction as growth direction of axons has a propensity to grow in approximately straight lines unless axons encounter obstructions or guidance cues (Sperry 1963; Easter et al. 1985; Yamamoto et al. 2002).

### 2.1.4 Proximity rule for establishing synapses

A synapse was established when the growing axon encountered another neuron within the connectible range (less than or equal to 1 unit length). The distance between a growth cone and a neuron was computed considering the radius of a neuron, or the shortest Euclidean distance between a growth cone and the surface of a neuron (the boundary of dendrites); thus, larger neurons increased their chances of establishing a synapse. We assumed that only the growth cone can establish a synapse.

### 2.1.5 Competition

Competition between neurons for synapse establishment was realized by limiting the number of incoming connections. A synapse was formed only when the neuron within vicinity could accommodate another synapse. All neurons keep growing their axons until they hit the border of the embedding space, as the maximum number of outgoing connections for a neuron was not limited assuming short-range connectivity within less than $700\,\mu m$.

### 2.1.6 Serial and parallel growth

For serial growth, each neuron takes turns to grow its axon, which represented no overlap in the time windows for axon growth; the first neuron grows out completely and forms all possible connections and finishes its growth, and only then the second neuron starts growing and makes synapses along the way. For parallel growth, all neurons start growing their axons simultaneously; thus, developmental time windows for all neurons coincided and maximally overlapped; all neurons start growing their axons at the same time, however, when the growth cone of a neuron hits the boundary of the space, that neuron stops axonal outgrowth. Thus, the finishing times for neurons are different (Fig. 1).

### 2.1.7 Developmental time windows

Different areas in the brain have shown dissimilar growth trajectories over time having partially overlapping time windows (Rakic 2002; Shaw et al. 2008; Sur and Leamey 2001). For instance, cortical neurons in Brodmann area (BA) 24 migrate to upper layers faster than neurons in BA11, BA46 and BA17; neurons in BA17 take the longest time to reach their final position (Rakic 2002). Moreover, previous studies have shown that neurons are inclined to establish synapses with other neurons whose time windows of growth overlapped (Kaiser and Hilgetag 2007; Nisbach and Kaiser 2007; Deguchi et al. 2011; Druckmann et al. 2014; Yu et al. 2009, 2012). Therefore, by comparing network features between serial and parallel growth, we could observe the influence of time windows for neuronal network development. Additionally, we tested partially overlapping time windows with a small partial overlap and a large partial overlap; serial growth is the extreme case of small overlap, i.e., zero overlap and parallel growth is the opposite end where time windows of axon growth are maximally overlapped (see details in Online material A10 and Figures A10–A12).

### 2.1.8 Data set

Positions of neurons and growth directions were generated constructing a total of 50 data sets to compare serial and parallel growth using equivalent positions of neurons and growth directions.

## 2.2 Comparison of growth scenarios

Serial growth and parallel growth were compared in terms of morphological, topological and spatial features. Morphological features included the number of established synapses, the number of potential synapses and the ratio between the two, or filling fraction (Stepanyants et al. 2002). In biological neuronal networks, not all potential synaptic locations are realized due to competition between neurons (Kaiser et al. 2009; van Ooyen et al. 2001; van Ooyen 2001), plasticity of connectivity or limitations in volume (Stepanyants et al. 2002). Next, topological properties such as out-degree, local efficiency and the proportion of bidirectional connections were investigated (Newman 2003; Brandes and Erlebach 2005; Costa et al. 2007). Out-degree of a neuron is the total number of outgoing connections from the neuron or the total number of outgoing synapses. Out-degrees of neurons were averaged over 50 trials for each neuron and ordered according to the sequence of serial growth. Then, this distribution was fitted with exponential or polynomial curves to assess the difference in out-degree as a function of the sequence of start. This shows whether earlier starters would have an advantage over later starters in establishing outgoing synapses. The maximum number of incoming connections was limited and increased according to the volume of a neuron. As a result, in-degree was constrained by the maximum number of incoming connections. Global efficiency is the inverse of the harmonic mean of the shortest path length between each pair of nodes (Eq. 1) and local efficiency for a node is calculated in the same way as global efficiency in the subgraph of the node comprised of its immediate neighbors (Eq. 2) (Latora and Marchiori 2001, 2003).

$$E_{\text{global}}(G) = \frac{1}{N(N-1)} \sum_{i \neq j} \frac{1}{L_{ij}} \qquad (1)$$

$$E_{\text{local}}(i) = E_{\text{global}}(G_i) \qquad (2)$$

where $N$ stands for the number of nodes in the network, $L_{ij}$ for the length of the shortest path between nodes $i$ and $j$, $G$ for a graph and $G_i$ for the subgraph that consists of neighbors of node $i$. The shortest path length of a pair of nodes is the length of the lowest number of edges to go from one node to the other node (Kaiser 2011).

Finally, we observed spatial properties of the grown neural connectivity. Connection probability between two neurons as a function of distance was calculated by dividing the number of connected edges by the number of possible connections given a distance between two neurons, where the distance between a pair of neurons was the Euclidean distance between the centers of the somata of neurons. Similarly, bidirectional connection probability as a function of distance was calculated by dividing the existing number of bidirectional connections by the number of all possible connections given a distance. The proportion of bidirectional connection was the ratio of bidirectional connections to all existing connections, and it was compared with that in the rewired networks to examine whether the proportion of bidirectional connections was higher in our random outgrowth model than that for random networks. We used rewired networks for benchmark random networks by randomizing or rewiring the preserving network while preserving degree distributions (Maslov and Sneppen 2002; Rubinov and Sporns 2010). The connection length between two connected neurons was the Euclidean distance between the centers of neurons assuming that the distance between the growth cone and the target neuron was negligible. If a neuron made multiple synapses until it hit the boundary of the given space, the locations where the intermediate synapses were established were considered as synaptic boutons and consequently the axon length for the neuron was defined as the distance between the neuron and the neuron where the last synapse was formed. Brain Connectivity Toolbox (Rubinov and Sporns 2010) was used to calculate network measures: in- and out-degree, local efficiency and the generation of random networks by rewiring our directed networks while preserving degree distributions.

## 2.3 Validation of our model prediction with *C. elegans* connectivity

We used data and information from previous studies based on the Worm Atlas (Chen et al. 2006; Varshney et al. 2011; Hall and Altun 2008) as we could make use of birth times of neurons in *C. elegans* as a proxy for developmental time windows (http://www.wormatlas.org/neuronalwiring.html#NeuronalconnectivityII). Unfortunately, there is no data from higher organisms since birth times or developmental time windows for axon growth and synaptogenesis are currently not available. For *C. elegans*, chemical synapses were considered, while electrical and neuromuscular junctions were excluded in the connectivity matrix. Spatial locations of neurons were obtained from a previous study (Choe et al. 2004). Based on birth times of neurons, we grouped neurons into three groups using k-means clustering: neurons in group 1 and group 2 have similar birth times and were born early, whereas neurons in group 3 were born much later compared to groups 1 and 2 (Fig. 5a). As each time k-means clustering provides slightly different clusters, we perform 50 trials and classified neurons with the most stable or frequent grouping. We assumed all neurons start to grow axons and dendrites after a similar latency period; birth time + $\alpha$ is the starting point of the time window of axon outgrowth. Thus, birth time can be directly associated with the start of the growth without losing generality since $\alpha$ is assumed to be about the same for all neurons. Groups are compared in terms of degrees, connection lengths and reciprocal connections. We tested the three most pronounced differences between serial and parallel growth: (1) whether earlier-born neurons acquired higher degrees than later-born neurons (group 1 vs. 2 and group 1 vs. 3), (2) whether earlier-born neurons established longer connections and (3) whether reciprocal connections are more numerous in neurons in groups 1 and 2 than for neurons in 1 and 3 or 2 and 3. Long-range connections were defined as connections where the length was at least one standard deviation above the mean of all connection lengths. Additionally, we examined local efficiency and connection probability as a function of distance (A13).

## 2.4 Statistical analysis

Topological and spatial properties between serial and parallel growth such as out-degree, local efficiency, connection probability and bidirectional connectivity were averaged over 50 trials for each neuron then fitted with exponential or polynomial curves. When curves followed close to a power law (connection probability and bidirectional connectivity), double-logarithmic axes were used and fitted with linear models. Higher proportion of bidirectional connectivity for parallel growth than serial growth was tested by paired t test and Wilcoxon signed rank test, two-tailed with an alpha level 0.05 and corrected by Bonferroni for multiple comparisons. To group neurons based on their birth times, we used $k$-means clustering using Euclidean distance. Degrees and long-range connection length were tested with Kruskal–Wallis test, and post hoc multiple comparisons were performed using Mann–Whitney test and corrected by Bonferroni. Calculations and statistical tests were performed with MATLAB R2012b (Mathworks Inc., Natick, MA). Algorithms are available online at http://www.dynamic-connectome.org/.

## 3 Results

### 3.1 Topological and spatial properties

#### 3.1.1 Serial versus parallel growth with a limit on the number of incoming connections

*Out-degree distribution* For serial growth, neurons that started growing their axons earlier took priority to establish synapses over late starters, hence a decreasing trend of out-degrees as the indices of order in development increased. In contrast, for parallel growth where every neuron started growing their axons simultaneously, out-degrees were independent of indices of neurons as the indices merely had nominal values in this case (Fig. 2a). The contrasting distributions of out-degrees for serial and parallel growth applied to all conditions independent of the number of neurons and the maximum number of incoming connections. However, the decreasing rate of out-degree for serial growth slowed down when allowing more incoming connections (less severe competition). Note that the less contrasting patterns between serial and parallel growth when larger numbers of incoming connections were allowed (e.g., the fourth column of Figure A4) should be attributed to milder competition among neurons rather than to higher neuronal density.

*Local efficiency* Neurons that started axon growth early on were also characterized by higher local efficiency compared to neurons that developed later for serial growth, whereas

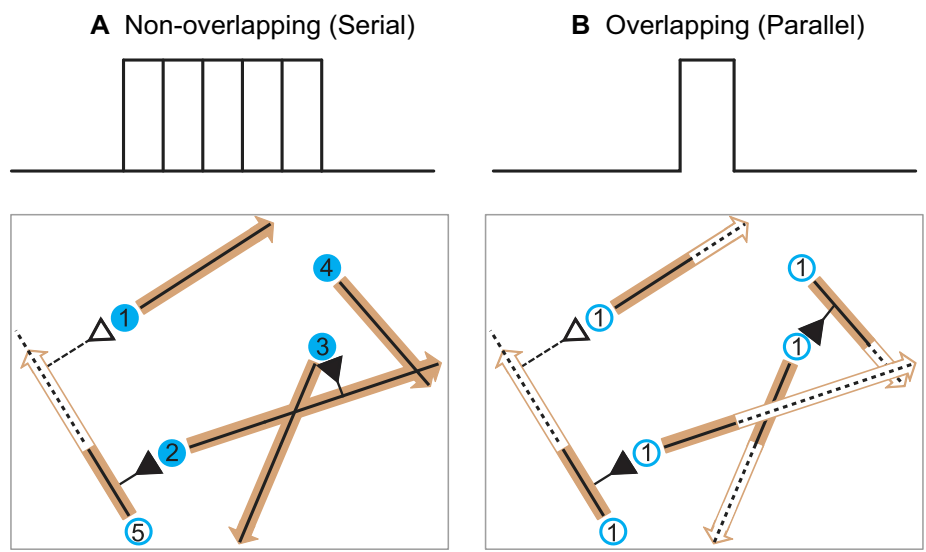

**Fig. 1** Simulation setting: serial versus parallel growth. Serial growth: neurons take turns to grow axons. Parallel growth: all neurons start to grow axons simultaneously. Continuous search mode: neurons examine all possible target neurons to establish synapses by finding whether the growth cone could find connectible neurons within a certain proximity, or whether the connectible space of growth cone intersects with the neuron sphere. *Blue circles*: *solid circles* denote neurons that finished axon growth and *empty circles* represent neurons that are active. *Numbers* in the *circles* represent the sequence of growth. *Black triangles* synapses, *black solid line* axons, *black dashed lines* future axon growth path

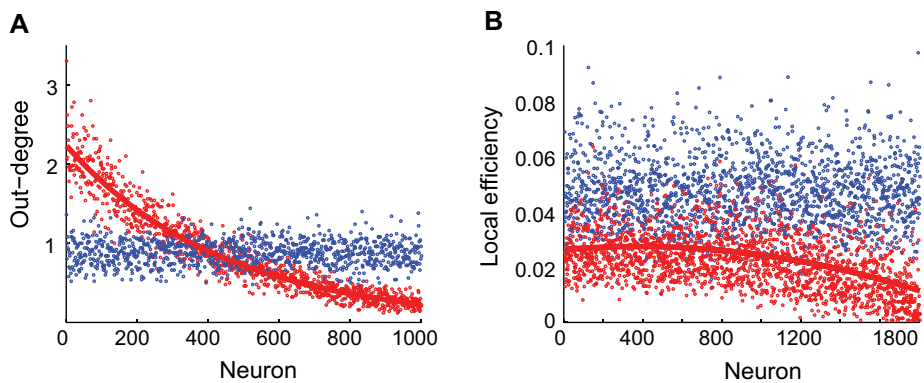

**Fig. 2** Topological properties for serial versus parallel growth. *x*-axis represents indices of neurons. For serial growth, the indices of neurons indicate the order of starting to grow axons, whereas for parallel growth the indices just represent nominal values. *Red* serial growth, *blue* par-

allel growth. **a** Out-degree (*y*-axis) 1,000 neurons with 0.5 radius and 1 maximum incoming connection, **b** local efficiency (*y*-axis) 1,800 neurons with 0.604 radius and 2 maximum incoming connections. For a complete overview of all conditions see Figures A4 & A5

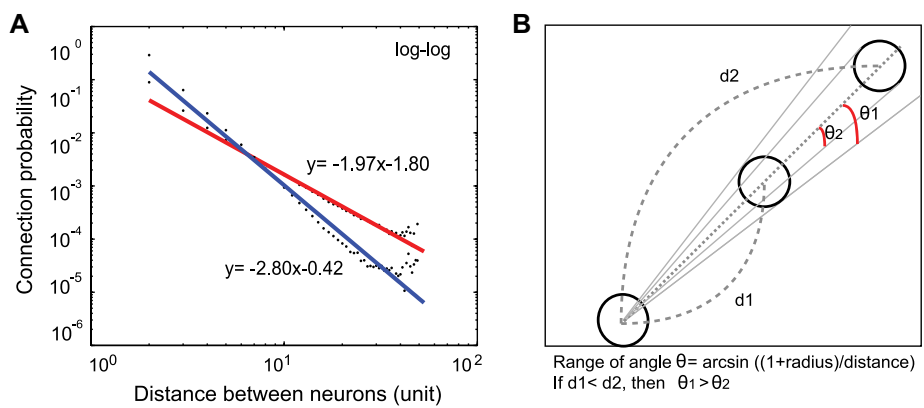

**Fig. 3** Connection probability and the schematic relationship between the range of angle for a presynaptic neuron can take and the distance between two neurons. **a** *Red* serial growth, *blue* parallel growth 1,400 neurons with radius 0.5 and 1 maximum incoming connection. For all conditions see Figure A6. **b** *Circles* neurons, *d*1 and *d*2: the distances between neurons, $\theta 1$ and $\theta 2$: the maximum direction angles a presy-

naptic neuron can have to establish synapses with a target neuron. As the distance becomes longer, the range of angles becomes narrower; if $d1 < d2$, then $\theta 1 > \theta 2$. Therefore, if a target neuron is far from the presynaptic neuron, it is less likely to form a connection, leading to lower connection probability there was no difference among neurons for parallel growth. The effect was not as pronounced as the decreasing trend for the out-degree distribution for serial growth; the decrease in local efficiency was observed mostly in later starting neurons for serial growth. Similar to the out-degree distribution, the disparate distributions of local efficiency for serial and parallel growth applied to all conditions with different numbers of neurons and maximum numbers of incoming connections; thus, we show a representative example (Fig. 2b). The decreasing rate of local efficiency for serial growth slowed down when allowing more incoming connections (Figure A5).

*Connection probability* Connection probability given a distance between two neurons decreased rapidly as the distance increased following power-law tail behavior (Figure A3). The connection probability decreased faster for parallel growth than for serial growth with distance between neurons, while the number of established synapses was the same between serial and parallel growth scenarios (Fig. 3a). The discrepancy of slopes for serial and parallel growth in the doubly logarithmic plot became reduced as the maximum number of incoming connections increased (Figure A6).

Figure 3b shows a schematic view of the relationship between the distance between neurons and the connection probability of the two neurons. When neurons are located farther apart from each other, the connection probability decreases since the range of growth directions toward which it can successfully establish a synapse is more limited. In other words, the connection probability between a pair of neurons located a distance *d* apart is proportional to the range of growth angles a neuron can take, that is the connection probability given a distance $d$, $P(d) \propto \theta$. The growth angle $\theta$ from the straight line between the centers of neurons can be calculated using the inverse sine as from our assumptions a neuron grows onwards, we can only consider $-\frac{\pi}{2} < \theta < \frac{\pi}{2}$.

**Fig. 4** Bidirectional connections and axon length. **a** *Boxplot* of the percentages of reciprocal connections for serial and parallel growth. **b** Axon length. Neurons that started axon growth earlier acquired longer axon lengths than later starters. *x*-axis for serial growth: the order of starting growth and *x*-axis for parallel growth: nominal indices of neurons, *y*-axis: axon length. For all conditions see Figure A7

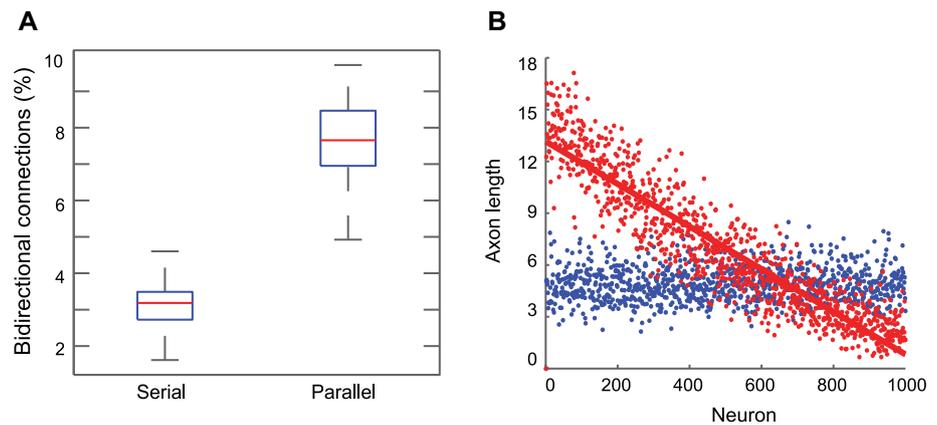

$$P(d) \propto \arcsin \frac{1+r}{d} \qquad (3)$$

$$\arcsin \frac{1+r}{d} \approx \frac{1+r}{d} \qquad (4)$$

$$P(d) \propto \frac{1}{d} \qquad (5)$$

where $r$ represents the size or the radius of a neuron and $d$ stands for the distance between the centers of neurons and inverse of sine can be approximated by its Maclaurin series when $1 + r < d$. Thus, Eq. 3 can be approximated as Eq. 4 taking only the first term of the series. The log–log plot of distance and angle also showed straight lines since the connection probability is inversely proportional to the distance $d$ (Eq. 5) (Figure A3).

*Bidirectional connections* Two neurons located close to each other are more likely to form synapses in general, and consequently, there would be more reciprocal connections between two neurons close to each other (Fig. 3b). However, this higher connection probability for bidirectional connections for nearby neurons assumes that relevant neurons are available for incoming connections. Synapses would not be formed when the target neurons have reached its maximum incoming limit even if neurons reside close to each other and the growth directions are narrow enough to form synapses. For serial growth even if both of them are within proximity of the other neuron's growth direction, one of them is more likely to be occupied and no longer available for making another synapse. Since they started at different time points, thus one neuron took much longer to reach the other neuron unlike for parallel growth. Therefore, bidirectional connections would be more numerous for parallel growth than for serial growth. As we expected, more frequent reciprocal connections were observed for parallel growth than for the serial growth scenario (Fig. 4a). For example, for 1,000 neurons with one incoming connection per neuron allowed, neurons formed about six times as many bidirectional connections for parallel growth as for serial growth (Wilcoxon signed rank test, $p < 10^{-7}$ corrected by Bonferroni). The difference between serial and parallel growth disappeared as neurons were allowed to have a large number of incoming connections (Figure A7). The simulated bidirectional connection probability using inverse sine (Eq. 5) was calculated by squaring the connection probability assuming independence among neurons for synapse establishment (Figure A3, black).

We also investigated whether the bidirectional connectivity was higher than expected in random networks. Random networks were constructed by randomizing or rewiring the original networks while preserving degree distributions (Maslov and Sneppen 2002; Rubinov and Sporns 2010). For smaller neuron sizes such as radii 0.5 and 0.604 concomitant with 1 and 2 maximum incoming connections allowed, respectively, rewired networks did not have any bidirectional connectivity even when considering larger numbers of neurons up to 1,800, or higher neuronal density. For larger neuron sizes, thus having a larger reach and more incoming connections allowed, originally generated networks with random outgrowth showed 10–17 times larger bidirectional connection proportions for serial growth and from 11 to about 40 times larger proportions for parallel growth depending on conditions. In summary, both serial and parallel growth resulted in higher bidirectionality than that of the benchmark random network, indicating that random outgrowth model with geometrical constraints can also reproduce more densely connected clusters.

*Connection length distribution and Axon length* The connection length distribution for parallel growth was characterized by an exponential decrease in the frequency, having a higher proportion of shorter connections, while the connection length distribution for serial growth demonstrated almost linear and slower decrease in the frequency having a larger number of longer connections than for parallel growth, while the total number of connections was the same both for serial

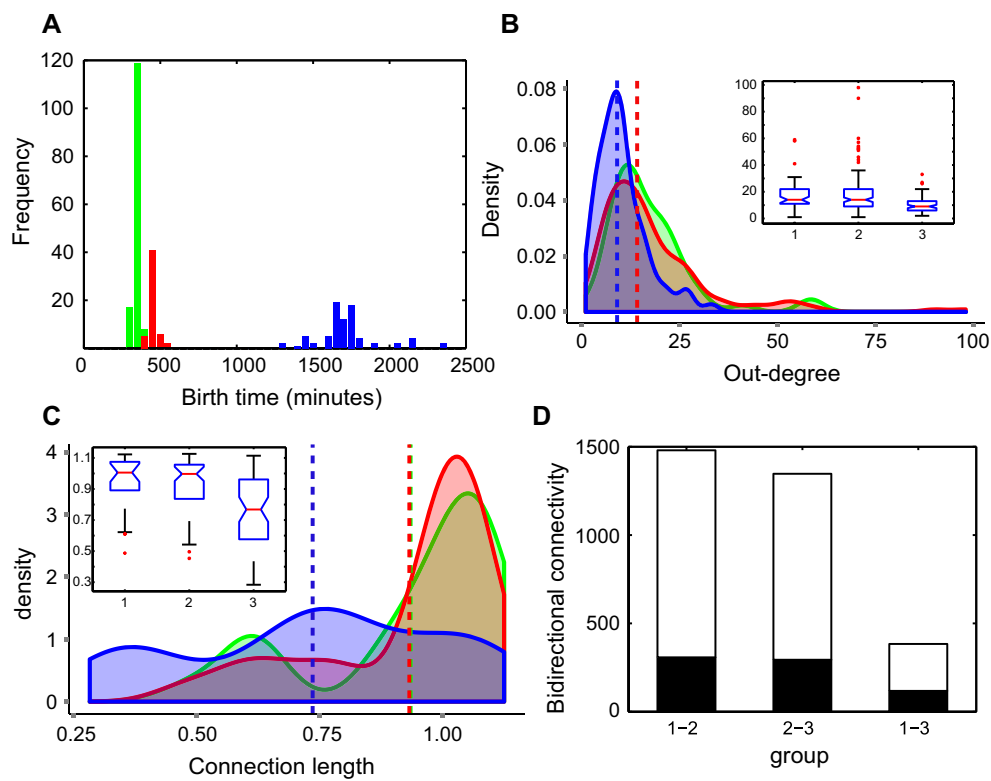

**A** Group 1, 2 and 3 using *k*-means clustering based on the birth times of neurons in *C. elegans*. *Green* group 1, *red* group 2, *blue* group 3. *x*-axis: birth time (minutes), *y*-axis: group membership. **b** Distributions of out-degree for group 1, 2 and 3 using kernel density estimation.

**Fig. 5** Model predictions versus neuronal connectivity in *C. elegans*. *x*-axis: out-degree, *y*-axis: density, *vertical dashed lines* medians of distributions, (*inset*) Birth time and degree. *Boxplot*. *x*-axis: birth time group, *y*-axis: degree of neurons. **C** Density plot of long-range connection lengths, *x*-axis: connection lengths, *y*-axis: density of distribution,

*vertical dashed lines* medians of connection lengths, (*inset*) Birth time and long-range connection lengths. *Boxplot* superimposed with data points. *x*-axis: birth time group, *y*-axis: length, approximated by Euclidean distance between centers of connected neurons, of long-range connections (mm). **d** Birth time group and the number of bidirectional connections. *x*-axis: birth time groups *y*-axis: the number of connections (*black* bidirectional connections; *white* the total number of connections between relevant groups) and parallel growth scenarios (Figure A8). Earlier starters demonstrated longer axon lengths than later starting neurons for serial growth, whereas no difference was observed in axon length for parallel growth (Fig. 4b).

### 3.2 Comparison of our model predictions with *C.elegans* connectivity

To validate our model predictions, we made use *C. elegans* data (Chen et al. 2006; Varshney et al. 2011; Hall and Altun 2008; Choe et al. 2004). Using only chemical synapses, we tested our three major predictions from our model concerning degree, connection lengths and bidirectional connectivity. Groups 1, 2 and 3 represent three groups of neurons clustered based on their birth times (Fig. 5a). While birth times in groups 1 and 2 were, on average, 114.78 min apart, the time difference between groups 2 and 3 was more than 1255.70 min. As birth times of group 1 and group 2 do not differ much, we can assume that group 1 and group 2 represent large overlapping time windows (or parallel growth), while group 1 and group 3 or group 2 and group 3 indicate small overlapping time windows (or serial growth).Here, birth time is used as an equivalent of the starting point of the time window for axon (and dendrite) outgrowth.[1]

We tested whether groups 1 and 2, with neurons born early during development, achieved higher out-degrees than group 3. We found that neurons in groups 1 and 2 indeed obtained larger numbers of connections (higher degrees) compared to neurons in group 3 (Kruskal–Wallis test, $p < 10^{-6}$ and post hoc multiple comparison Mann–Whitney two-tailed: between groups 1 and 3, $p < 10^{-4}$; between groups 2 and 3, $p < 10^{-4}$, $p$ values are corrected by Bonferroni, Fig. 5b). Neurons in groups 1 and 2 established longer connections compared to neurons in group 3 (Kruskal–Wallis test, $p < 10^{-4}$; Mann–Whitney two-tailed: between groups 1 and 3, $p < 0.0057$; between groups 2 and 3, $p < 0.0002$, $p$ values are corrected by Bonferroni, Fig. 5c) and more bidirectional connections (Fig. 5d). As expected for parallel growth, group 1 versus group 2 did not display significantly higher degree, long-range connection lengths or bidirectional connectivity differences. Local efficiency and connection probability, however, were not consistent with the model predictions indicating that there are factors excluded in the model which influence these features (See Discussion and A13).

---

[1] Time windows for neurogenesis and synaptogenesis should be treated differently; however, here we assumed time windows for synaptogenesis in *C. elegans* started after equivalent time passes for all neurons for simplicity.

## 4 Discussion

In this study, we demonstrated that different time windows for axon growth could lead to distinct topological and spatial characteristics by exploring two extreme cases of time windows representing serial (heterogeneous) and parallel (homogeneous) growth. We also tested our model predictions with *C. elegans* connectivity data. Overlapping and non-overlapping time windows for axon growth resulted in different topological and spatial properties of neuronal networks, although morphological properties such as the number of potential synapses and established synapses were not different (A2). For serial growth, neurons that started axon growth early on achieved higher out-degrees, higher local efficiency and longer axon lengths than later starting neurons, while no difference was observed for parallel growth. Bidirectional connections were more numerous for parallel growth. Finally, axon lengths for serial growth were longer. Together, time windows for axon growth seem to have a major influence on network organization during neural development.

### 4.1 Non-overlapping versus overlapping time windows for axon growth

Brain development shows region-specific time windows of growth that partially overlap (Rakic 2002; Shaw et al. 2008; Sur and Leamey 2001). Previous computational studies (Kaiser and Hilgetag 2007; Nisbach and Kaiser 2007) as well as the analysis of the *C. elegans* cell lineage (Varier and Kaiser 2011) strongly suggest that neurons are inclined to establish synapses with other neurons whose developmental time windows overlap (Varier and Kaiser 2011; Deguchi et al. 2011; Druckmann et al. 2014). We investigated how developmental time windows for axon outgrowth affect network connectivity by analyzing non-overlapping (serial growth) and maximally overlapping (parallel growth) time windows for axon growth. For serial growth, a neuron is able to start growing its axon only after the previous neuron finishes developing its axon. Thus, time windows of axonal growth do not overlap, whereas for parallel growth, all neurons have the same time window onset, starting to grow axons simultaneously. Note, however, that the end point of the time window—the time when a neuron left the embedding space—could differ between neurons. Biologically, neurons with highly overlapping time windows of development can be interpreted as neurons whose birth times, lineage and cell types are homogeneous such as cortical neurons in the same layer or clone sister neurons sharing genetic resemblance, both of which were characterized with a higher propensity to establish synapses between them (Li et al. 2009; Deguchi et al. 2011).

For serial growth, earlier development for axon growth facilitated more numerous synapse establishments resulting in higher out-degrees, higher local efficiency and longer axon lengths, suggesting a possible mechanism for network hub formation. These results are in line with experimental findings analyzing the neuronal network of *C. elegans* (Varier and Kaiser 2011) and even with findings for the network of fiber tracts between cortical regions in the macaque (Kaiser and Varier 2011). As we expected, earlier-born neurons in *C. elegans* established more connections than later-born neurons. This might allow us to predict the history of neural development based on cell lineage and adult degree distribution. Local efficiency shows that how efficiently neighbor neurons of a neuron would communicate when the neuron is removed being linked to the fault tolerance of the network (Latora and Marchiori 2001, 2003). Here, higher local efficiency of early starter neurons indicates that the local network comprised of the neuron's immediate neighbor neurons has more efficient communication among neighbor neurons and also more resilience against the removal of the early starter neuron. In line with these findings, an earlier study in *C. elegans* (Varier and Kaiser 2011) found that most connected neurons were born at similar time points and that the majority of long-distance connections appeared early on. This suggests that overlapping developmental time windows could contribute to increase the connection probability and early establishment of long-distance connectivity, which could secure specifically targeted long-range connections. Starting early on is a mechanism for individual neurons to establish long-distance connections. However, serial growth also affected the neural population as a whole. In our simulations, the sequential serial growth generated significantly more long-distance connections than the more homogeneous parallel growth (Fig. 4b); neurons that started axon growth later often found that postsynaptic neurons were already occupied, whereas earlier starters successfully established synapses to the same target neurons even if they were distantly located.

The connection probability between a pair of neurons decreased as the distance between them increased in line with anatomical studies (Hellwig 2000; Schuz 2001; Kaiser et al. 2009). The rate of decrease was steeper for parallel growth indicating that neurons with the same time window of axonal growth tend to prefer short-distance connections rather than long-distance connections. In contrast, for serial growth, the distance was not the only factor to establish connections since late starter neurons may not be able to form certain synapses due to the limited number of incoming connections, thus having a slower decrease in connection probability with distance.

Bidirectional connections were more numerous for parallel growth than for serial growth, which we confirmed with *C. elegans* connectivity data; the discrepancy between neurons (or neuron groups) was negatively correlated to the degree of overlap in the developmental time windows. In other words, larger overlap of developmental time windows (groups 1 and 2) reduced differences in degrees and connection length. More frequent reciprocal connections for parallel growth provide additional converging evidence that overlapping time windows during development would produce more reciprocal connections between neurons (Kaiser and Hilgetag 2007; Nisbach and Kaiser 2007; Varier and Kaiser 2011). The bidirectional connection probability decreased more rapidly than the overall connection probability (Figure A3), which is consistent with previous studies using thick-tufted layer 5 pyramidal neurons in neonatal Wistar rats (Perin et al. 2011, 2013). Previous studies have shown overrepresented reciprocal connections in the rat relative to random networks claiming that the synaptic connectivity is preferential rather than random (Kelsch et al. 2010; Markram et al. 1997). However, in this study, we observed a higher proportion of bidirectional connections while still using a random outgrowth mechanism for both serial and parallel growth scenarios.

Earlier-born neurons in *C. elegans* acquired higher outdegree, longer axon lengths and higher reciprocal connectivity, which were consistent with the model predictions. However, local efficiency and connection probability as a function of distance between neurons showed discrepancy from what the model predicted (A13). We believe that the model predictions and the actual results from *C. elegans* were different because (1) differences in local efficiency between serial and parallel growth were less apparent for all conditions (Figure A5) and (2) connection probability in our model depends mainly on the geometrical arrangement of dendrites (neuron spheres) and axons, whereas the connectivity of *C. elegans* has additional constraints such as its elongated body shape and a higher prevalence of long-distance connections (Kaiser and Hilgetag 2006).

## 4.2 Limitations and future studies

This general study of axon growth uses simplifications both for axon growth and neuron morphology. The size of the neuron and the proximity rule can only be an estimate of the average behavior of axons growing close to existing neurons.

For models of specific tissue, the morphology of the dendritic tree and the number of spines would need to be taken into account. Such parameters for many types of neurons and many different species are available in the NeuroMorpho database (Samsonovich and Ascoli 2003; Ropireddy and Ascoli 2011; Zawadzki et al. 2012). Another simplification is the axon growth in a straight line. Even though growing in a straight line is the default behavior, axons can branch or their growth directions can be influenced by attractive or repulsive signaling cues in the external environment (Yamamoto et al. 2002; Krottje and Van Ooyen 2007; Sakai and Kaprielian 2012). Finally, the embedding space of neurons for axon growth and synaptogenesis of our model was fixed during development, while internal volume changes through neurite growth and external mechanical factors could change the location of neurons and influence their synapse formation probabilities. For uniform expansion along all directions, this would increase connection lengths, but differences between serial and parallel growth would remain (See A14 for detailed analysis and discussion).

## 4.3 Conclusion

In the current study, we showed that for serial growth of axons, neurons with an early start of axonal growth acquired higher out-degrees, higher local efficiency and longer axon lengths, while overlapping time windows for parallel growth contributed to higher reciprocal connection and faster decrease in overall connection probability and connection length distribution with an increased distance between neurons. These predictions were confirmed when comparing our findings with the organization and development of the neuronal network of *C. elegans*. In summary, we demonstrated that axon growth time windows—like time windows for synaptogenesis and neuronal migration—modulate the topological and spatial properties of neuronal networks. We hope that these findings elucidate the origins of normal and pathological network development.

**Acknowledgments** S.L. and M.K. were funded by National Research Foundation of Korea funded by the Ministry of Education, Science and Technology (R32-10142). M.K. and S.L. were also funded through the Human Green Brain Project (http://www.greenbrainproject.org/) supported by EPSRC (EP/ K026992/1) and (EP/G03950X/1).